\documentclass{pasj00}
\draft
\usepackage{colordvi}
\begin{document}
\SetRunningHead{T.Anan et al.}{Spicule Dynamics over Plage Region}
\Received{2009/7/11}
\Accepted{2009/10/15}

\title{Spicule Dynamics over Plage Region}

\author{
Tetsu \textsc{Anan}\altaffilmark{1,2},
Reizaburo \textsc{Kitai}\altaffilmark{2,3},
Tomoko \textsc{Kawate}\altaffilmark{1,2},
Takuma \textsc{Matsumoto}\altaffilmark{1,2},
Kiyoshi \textsc{Ichimoto}\altaffilmark{2},
Kazunari \textsc{Shibata}\altaffilmark{2,3},
Andrew \textsc{Hillier}\altaffilmark{1,2},
Kenichi \textsc{Otsuji}\altaffilmark{1,2},
Hiroko \textsc{Watanabe}\altaffilmark{1,2},
Satoru \textsc{UeNo}\altaffilmark{2},
Shin'ichi \textsc{Nagata}\altaffilmark{2},
Takako \textsc{T.Ishii}\altaffilmark{2},
Hiroyuki \textsc{Komori}\altaffilmark{2},
Keisuke \textsc{Nishida}\altaffilmark{2},
Tahei \textsc{Nakamura}\altaffilmark{1,2},
Hiroaki \textsc{Isobe}\altaffilmark{3},
Masaoki \textsc{Hagino}\altaffilmark{4}
 } 

\altaffiltext{1}{Department of Astronomy, Kyoto University, Kitashirakawa-Oiwake-cho, Sakyo-ku, Kyoto 606-8502} 
\altaffiltext{2}{ Kwasan and Hida Observatory, Kyoto University, Kamitakara, Gifu 506-1314, Japan }
\altaffiltext{3}{ Unit of Synergetic Studies for Space, Kyoto University, Japan} 
\altaffiltext{4}{ National Astronomical Observatory of Japan, Mitaka, Japan }

\email{anan@kwasan.kyoto-u.ac.jp}


\maketitle

\begin{abstract}

We studied spicular jets over a plage area and derived their dynamic characteristics using Hinode Solar Optical Telescope (SOT) high-resolution images. 
The target plage region was near the west limb of the solar disk.
This location permitted us to study the dynamics of  spicular jets without the overlapping effect of spicular structures along the line of sight.

In this work, to increase the ease with which we can identify spicules on the disk, we applied the image processing method `MadMax' developed by Koutchmy et al. (1989).
It enhances fine, slender structures (like jets), over a diffuse background. 
We identified 169 spicules over the target plage.
This sample permits us to derive statistically reliable results regarding spicular dynamics.

The properties of plage spicules can be summarized as follows:
(1) In a plage area, we clearly identified spicular jet features. (2) They were shorter in length than the quiet region limb spicules, and followed ballistic motion under constant deceleration. (3) The majority (80\%) of the plage spicules showed the cycle of rise and retreat, while 10\% of them faded out without a complete retreat phase. (4) The deceleration of the spicule was proportional to the velocity of ejection ( i.e. the initial velocity ).

\end{abstract}


\section{Introduction}

Spicules are one of the most fundamental elements in the solar chromosphere.
According to the review by Michard (1974) they are very thin spike-like features, show rapid temporal change and are observed ubiquitously at the solar limb.
However, it has been difficult to fully follow the spicule evolution and to understand their physics due to limits in the spatiotemporal resolution of previous observations (Sterling 2000).

In the past, it had been reported that spicules are absent over plage regions, where the magnetic field is relatively strong.
Spicules have only been found in plagettes (the small plages of the network) or at the edges of larger plage regions (Zirin 1974).
Based on the shock-driven model of quiet region spicules by Suematsu et al. (1982), Shibata \& Suematsu (1982) suggested that the lengths of spicules in the plage atmosphere are smaller than those in quiet atmosphere and so they could not be observed.

Recent developments in the manufacture of observing instruments, detectors and in image processing, has enabled us to study spiculer jets over active regions (usually called ``Dynamic Fibrils'') and limb spicules in detail.
De Pontieu et al. (2004, 2007a) discussed spicular jets over plage regions and argued that the same features were seen as `Type-\emissiontype{I}' spicules on the limb.
Moreover it is reported that there is a close relationship between these spicular jets (Dynamic Fibrils) and limb spicules (Christopoulou et al. 2001; Hansteen et al. 2006; De Pontieu et al, 2007b; Rouppe van Voort et al. 2007; Zaqarashivili et al. 2007).
In addition, there are many observations in H$\alpha$ showing upper chromospheric oscillations above active region plage.
These oscillations are believed to correspond with periodic flows in spicular structure (De Pontieu et al. 2003).
De Pontieu et al. (2007a) also reported significant differences of fibril properties between a dense plage region and a thin plage region surrounding two small sunspots of NOAA AR 10813.
However, systematic study of spicular dynamics over plages has not yet been fully performed.

Our motivation of this study is to verify the existence of spicular jets over plages and derive their dynamic characteristics using Hinode Solar Optical Telescope (SOT) high-resolution images.
The target plage region was near the west limb of the solar disk.
This location permits us to study the dynamics of spicular jets without the problems caused by the overlapping effect of other spicular structures along the line of sight.

We applied the image processing method `MadMax' (Koutchmy et al. 1989) to the Ca\emissiontype{II} H images observed with the Hinode/SOT Broadband Filter Imager(BFI), and identified 169 spicules over the plage region.
Of them, 147 spicules showed a parabolic trajectory, similar to quiet region H$\alpha$ mottles reported by Suematsu (1998).
This sample of 147 spicules permits us to derive statistically reliable results on spicular dynamics.
 
In the following sections, we describe the details of the observations and the image processing ({\S}2), the method of analysis for this data ({\S}3), the results of basic dynamical parameters for spicules over the plage ({\S}4), and finally discuss and summarize our findings ({\S}5).


\section{Observation and Image Processing}

\begin{figure}
  \begin{center}
    \FigureFile(160mm,80mm){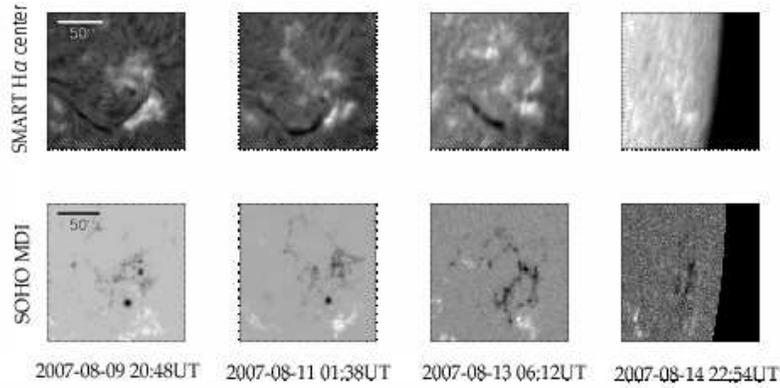}
  \end{center}
  \caption{
  Evolution over a 5 day period of the target region.
  Top row shows the variation in H$\alpha$ center observed with SMART at the Hida observatory of Kyoto University, and the bottom row shows the magnetic field in the photosphere observed with the MDI on board SOHO.
  From the left, the observed dates are August 9, August 11, August 13 and August 14, respectively.
  
  }\label{fig:periplot}
\end{figure}

\begin{figure}
  \begin{center}
    \FigureFile(160mm,80mm){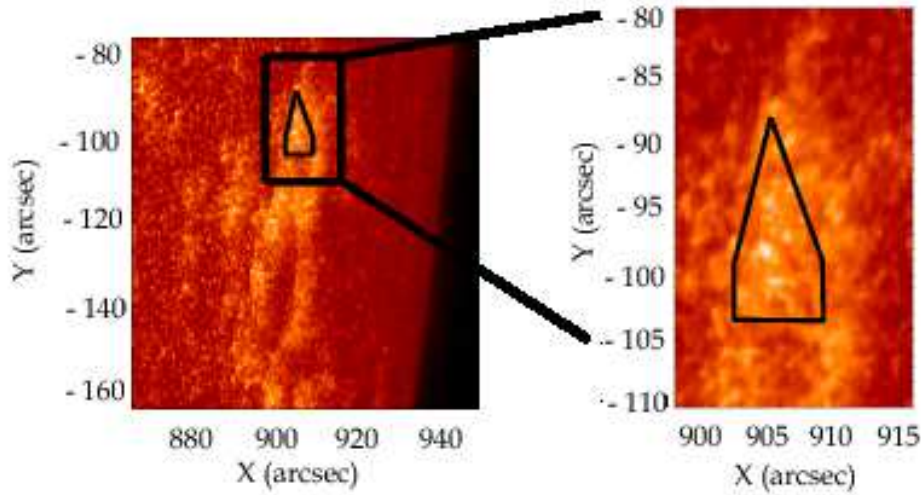}
  \end{center}
  \caption{
  Two Ca\emissiontype{II} H images with Hinode/SOT Broadband Filter Imager for plages on 2007 August 14.
  The right image is the enlargement of the region indicated by the black box on the left image. 
  The analyzed area is shown by thin black lines on the right Ca\emissiontype{II} H image.
  }\label{fig:periplot}
\end{figure}

Figure 1 shows the evolution of the target plage region over a period of 5 days. 
The evolution in H$\alpha$ center was observed with the Solar Magnetic Activity Research Telescope (SMART) at Hida observatory of Kyoto University.
The pixel size and the time resolution were $\timeform{0".5}$ and 7 sec, respectively.  
The magnetic field in the photosphere was observed with the Michelson Doppler Imager (MDI) (Scherrer et al. 1995) on board SOHO, with pixel size and time resolution of $\timeform{2"}$ and 60 sec respectively.
The target plage region was a remnant of a bipolar active region.

We use a series of images observed with the Hinode/SOT BFI in Ca\emissiontype{II} H 3968{\AA} wavelength. The observed intensity through the Ca\emissiontype{II} H filter of 3{\AA} pass band is a combination of photospheric and chromospheric radiation. The observation was made from 22:05 UT to 23:59 UT on 2007 August 14 with a fixid cadence of 45 sec and an exposure time of 0.15 sec. 
The spatial pixel size was $\timeform{0".109}$.

The target plage region was near the west solar limb on 2007 August 14 (figure 2). The heliocentric coordinate of the plage region at the time observation was (S\timeform{8D}, W\timeform{72D}).
The IDL routine {\it fg\_prep.pro}, which is part of the Hinode tree of solarsoft, was applied for the purpose of dark subtraction and flat fielding.

In order to enhance the contrast between the spicules and the background, we apply the image processing routine `MadMax', a directionally sensitive operator.
For each pixel the maximum of the second derivatives in the 8 spatial directions around the pixel is computed (Christopoulou et al. 2001).
When plotted as an image, this method highlights faint slender structures, allowing us to pursue the temporal evolution of each spicule.

Figure 3 shows the temporal evolution of a spicule at 45 sec intervals.
The figure shows that a faint, slender structure in the Ca\emissiontype{II} H images is enhanced by applying `MadMax'.
We followed the temporal evolution of these enhanced features, and identified these features as spicules when they clearly showed rise and retreat motion or only rise motion for more than 3 frames.  
We used eye pointing to mark the locations of the top and bottom of the spicules, and identified 169 spicules in this region.   

\begin{figure}
  \begin{center}
    \FigureFile(80mm,80mm){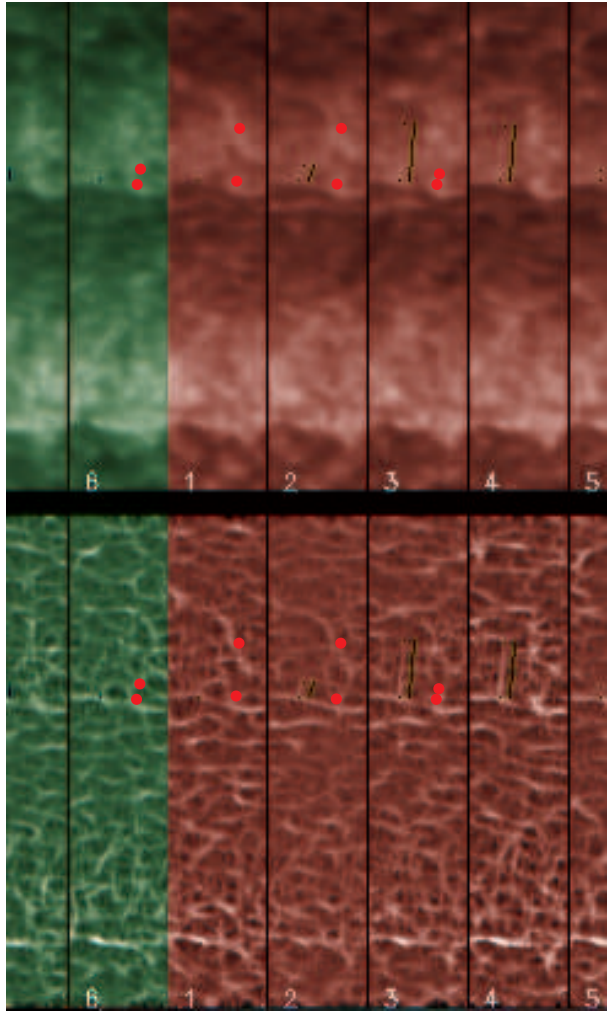}
  \end{center}
  \caption{
 	Top row shows the time series of the 6 images of Hinode/SOT Broadband Filter Imager in the Ca\emissiontype{II} H.
	Bottom row shows the 6 images processed by `MadMax'.
	West limb is in upward derection.
	In each sub-images, we mark two red points which show the top and footpoint of a spicule, and a red line along the spicule length on its right side. 
  }\label{fig:periplot}
\end{figure}


\section{Method of Analysis}
To derive the characteristics of the jet dynamics, we tried to fit the time-height curve with a parabola using the least square method. 
When we try to fit a parabola to the time-height relation of a short duration spicule, one which was only detected in three consecutive frames, the derived parabola parameters are influenced heavily by the measurement errors.
Generally speaking, the more data points used for the fitting, the greater accuracy in the fitting.
To improve the accuracy, we tried to increase the number of data points for fitting by including the start-time data of the spicule ejection.
The start time of the jet was unknown but it had to be in the 45 sec time interval prior to the first detection of the jet.
We divided the previous 45 sec into 20$\times$2.25 sec intervals. 
By assuming that the start time of jet is in one of these intervals, we can fit a time-height curve for every assumed start time. 
By selecting the solution with the least square, we could derived the best fit solution. 
An example of parabola fitting to a `parabolic' spicule is shown in figure 4.

After applying this method to all the measured jets, we classified the solutions as follows; 
if the predicted end time, which is defined as the time when the fitting parabola reaches zero, was within 45 sec after the time of final detection, we classified the jet as `parabolic' (figure 4). 
Conversely, when the predicted end time was not within 45 sec time after the final detection time, we classified it as `fade out' (figure 5). 
The time-height plots of the jets whose fitted solutions exhibited large mean square deviation from the observation were found to be very complex, and were classified as `irregular'. 

 We derived the dynamical parameters of jets (maximum length, life time, maximum velocity and acceleration) from the fitted parabola function for both the `parabolic' and the `fade out' types. 
For the `fade out' jets, the lifetime was defined as the time span from the fitted start time to the final detection time plus 22.5 sec, as the real end time was uncertain due to the observational cadence time of 45 sec.

 A point of note will be given here with regard to the effect of geometrical projection. 
Correction of the projection effect was not taken into account in this work, due to the lack data regarding the tilt angle of spicules from the normal direction.

\begin{figure}
  \begin{center}
    \FigureFile(80mm,80mm){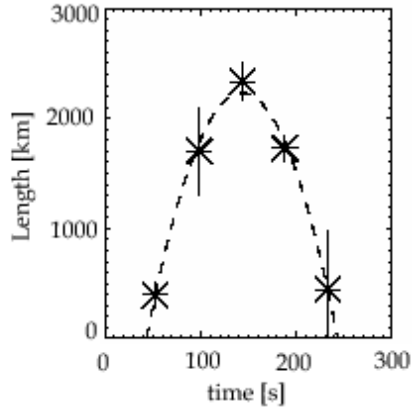}
  \end{center}
  \caption{
  The time evolution of a `parabolic' spicule length.
  The asteriskes are measured data points. 
  The dashed line indicates the fitted parabola line.       
  }\label{fig:periplot}
\end{figure}

\begin{figure}
  \begin{center}
    \FigureFile(80mm,80mm){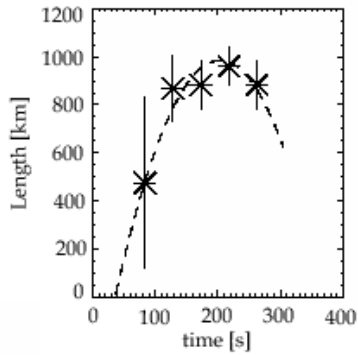}
  \end{center}
  \caption{
  The time evolution of a `fade out' spicule length.
  The others are the same as figure 4.
  }\label{fig:periplot}
\end{figure}


\section{Results}

We identified 169 spicules over the plage area from the Hinode Ca\emissiontype{II} H images.
Among them, $125$ spicules were `parabolic', $22$ spicules were `fade out', and the remaining 22 spicules were `irregular'.
From this we can say that most spicules in plage regions show parabolic trajectories, similar to quiet region disk spicules as reported by Suematsu (1998).   
In figure 6 and 7, we show histograms of the dynamic parameters for the `parabolic' type and the `fade out' type spicules, respectively.
The following is an individual study of each parameter.
 
\subsection{Maximum Length}

The maximum length of a spicule over a plage region was found to be significantly shorter than that of a typical quiet limb spicule.
The mean heights of the `parabolic' and the `fade out' spicules are 1300\,km and 820\,km respectively.
In contrast, the maximum length of a quiet limb spicule is between 6500\,km $\sim$ 9500\,km (Beckers 1968, 1972).
Recently, De Pontieu et al. (2007b) reported the existence of `Type-\emissiontype{II}' spicules which show rapid rise and abrupt fading at the end of their cycle, similar to the `fade out' spicules found in this paper.
The `Type-\emissiontype{II}' spicules were found to have a maximum length between 1000\,km $\sim$ 7000\,km.     
From this we can conclude that spicules over plage are significantly shorter than those found in quiet regions.
The result is consistent with the result of the shock-driven model of spicules by Shibata \& Suematsu (1982).
They claimed that spicules over plage area are very short and as a result are hard to detect.
Among the plage spicules, the `parabolic' spicules are generally longer than `fade out' spicules. 

\subsection{Life Time}

As was stated in section 2, we disregarded those spicules with a lifetime less than 90 sec (three consecutive frames in 45 sec cadence observation).
Under that restriction, the mean lifetimes of the `parabolic' and the `fade out' spicules are 179 sec and 197 sec respectively.
By contrast, life times of the quiet region limb spicules are 60 sec $\sim$ 600 sec (Lippincott 1957; Bray \& Loughhead, 1974).
We can see from figure 6 and 7 that the lifetimes of the `parabolic' and the `fade out' spicules have a similar distribution.
This allows us to conclude that plage spicules have shorter lifetimes than those of quiet region spicules.

\subsection{Maximum Velocity}

The mean maximum velocities of the `parabolic' spicules, the `fade out' spicules and the typical quiet region limb spicules are 34.4\,km s$^{-1}$, 15.9\,km s$^{-1}$ and 25\,km s$^{-1}$, respectively.
The maximum velocity range of the `parabolic' spicules is between 6\,km s$^{-1}$ $\sim$ 110\,km s$^{-1}$, while that for typical quiet region limb spicules is  10\,km s$^{-1}$ $\sim$ 50\,km s$^{-1}$ (Suematsu 1998), whereas it is 5\,km s$^{-1}$ $\sim$ 40\,km s$^{-1}$ for the `fade out' spicules.

In figure 8, we show the scatter plot of the maximum length ($L_{max}$) vs. the maximum velocity ($V_{max}$), with black diamonds denoting the `parabolic' spicules and blue crosses denoting the `fade out' spicules.
It should be noted that data points of the short duration spicules are not plotted in figure 8, as we excluded from our analysis such spicules that were detected only in one or two consecutive frames. 
When an object is ejected from a height 0 with an initial speed of $V_{max}$ at $t$=0 under the deceleration of $d$, its speed $V$ will change as $V=V_{max}-dt$, this is known as ballistic motion. 
Its height $h$ is given by $h=V_{max}t-d\,t^2/2$. 
The maximum height ($L_{max}$) is expressed as $L_{max}=V_{max}^{2}/2d$.
Thus the lifetime of the ballistic motion ($T_{l}$) is given by $T_{l}=2V_{max}/d=4 L_{max}/V_{max}$.  
As the lifetime $T_{l}$ of a short duration spicule, which was detected only in one frame or in two consecutive frames of 45 sec cadence, will be shorter than the 2$\times$45 sec, the data points of the short duration spicules would be plotted below the dotted line $L_{max}/V_{max}=2\times45/4$ in figure 8, if they were included in our analysis.

The data points show a positive correlation between the maximum length and the maximum velocity.
The `fade out' spicules are found along the short length-slow speed part of the correlation.
To interpret the positive correlation, we refer to the simulation results for plage region spicule performed by Shibata \& Suematsu (1982).
Their model is a shock-driven model of spicules excited by a sudden pressure enhancements in the low atmosphere. 
Their simulation results are plotted in figure 8 with red asterisks.
The maximum lengths of the simulated spicules are larger than those observed.
We think that this difference in the maximum length can be explained as follows.
According to their analysis, the following relations hold for shock-driven spicules 
\begin{eqnarray*}
&&  maximum\,\,length \propto (\rho_{h_o}/\rho_{c})^{0.46},\\
&&  maximum\,\, velocity \propto (\rho_{h_o}/\rho_{c})^{0.23},
\end{eqnarray*}
where $\rho_{h_o}$ is the density at which the pressure enhancement occurs and $\rho_{c}$ is the coronal density.  
According to these equations, the maximum length exhibits a greater dependence on $\rho_{h_o}$ than the maximum velocity does. 
If we assume lower values of the $\rho_{h_o}$ than the Shibata \& Suematsu (1982) values, we expect that the reduction in the maximum length will be greater than the reduction in the maximum velocity.
Then, the theoretical prediction will match the observed correlation.

\subsection{Acceleration}

The mean acceleration for the `parabolic' spicules and the `fade out' spicules are -0.51\,km s$^{-2}$ and -0.13\,km s$^{-2}$ respectively.
Figure 7 shows that the deceleration of the `fade out' spicules is less than deceleration due to solar gravity.
Figure 6 shows that the deceleration of the `parabolic' spicules ranges from above to below the value for deceleration due to solar gravity.

In figure 9, we show a scatter plot of the deceleration vs. the maximum velocity, with black diamonds denoting the  `parabolic' spicules and with blue crosses denoting the `fade out' spicules.
As was discussed in subsection 4.3, we excluded from our analysis spicules that were detected only in one or two consecutive frames. 
The data points of short duration spicules would have been plotted below the dotted line ($V_{max}/d=2\times45/2$sec) in figure 9. 

The observed data points follow a linear positive correlation, which means that a plage spicule decelerates heavily when its maximum speed is high.
The linear positive relation in figure 9 is consistent with the result of the shock-driven model simulated by Heggland et al. (2007).
In figure 9 the `fade out' spicules are distributed in the slow speed-weak deceleration region.

\begin{figure}
  \begin{center}
    \FigureFile(80mm,80mm){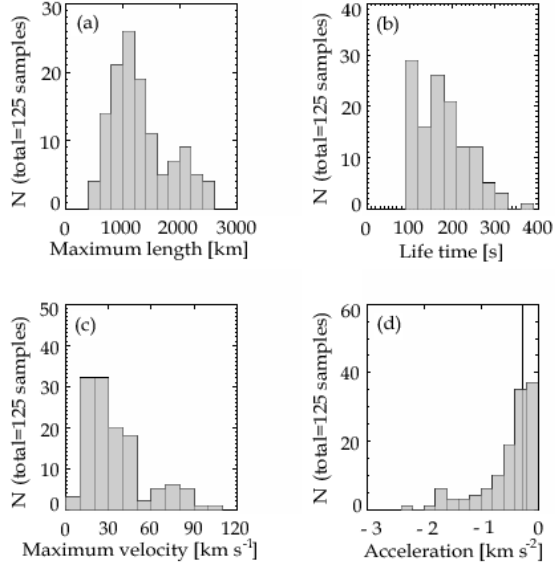}  
  \end{center}
  \caption{
  Histogram of (a) the maximum length, (b) the life time, (c) the maximum velocity and (d) the acceleration of 125 `parabolic' spicules.
  The mean value of each spicule's maximum length, life time, maximum velocity and acceleration is 1300\,km, 179\,s, 34.4\,km s$^{-1}$ and $-$0.51\,km s$^{-2}$, respectively.
  The solid line in the histogram of acceleration indicate the solar surface gravity.     
  }\label{fig:periplot}
\end{figure}

\begin{figure}
  \begin{center}
    \FigureFile(80mm,80mm){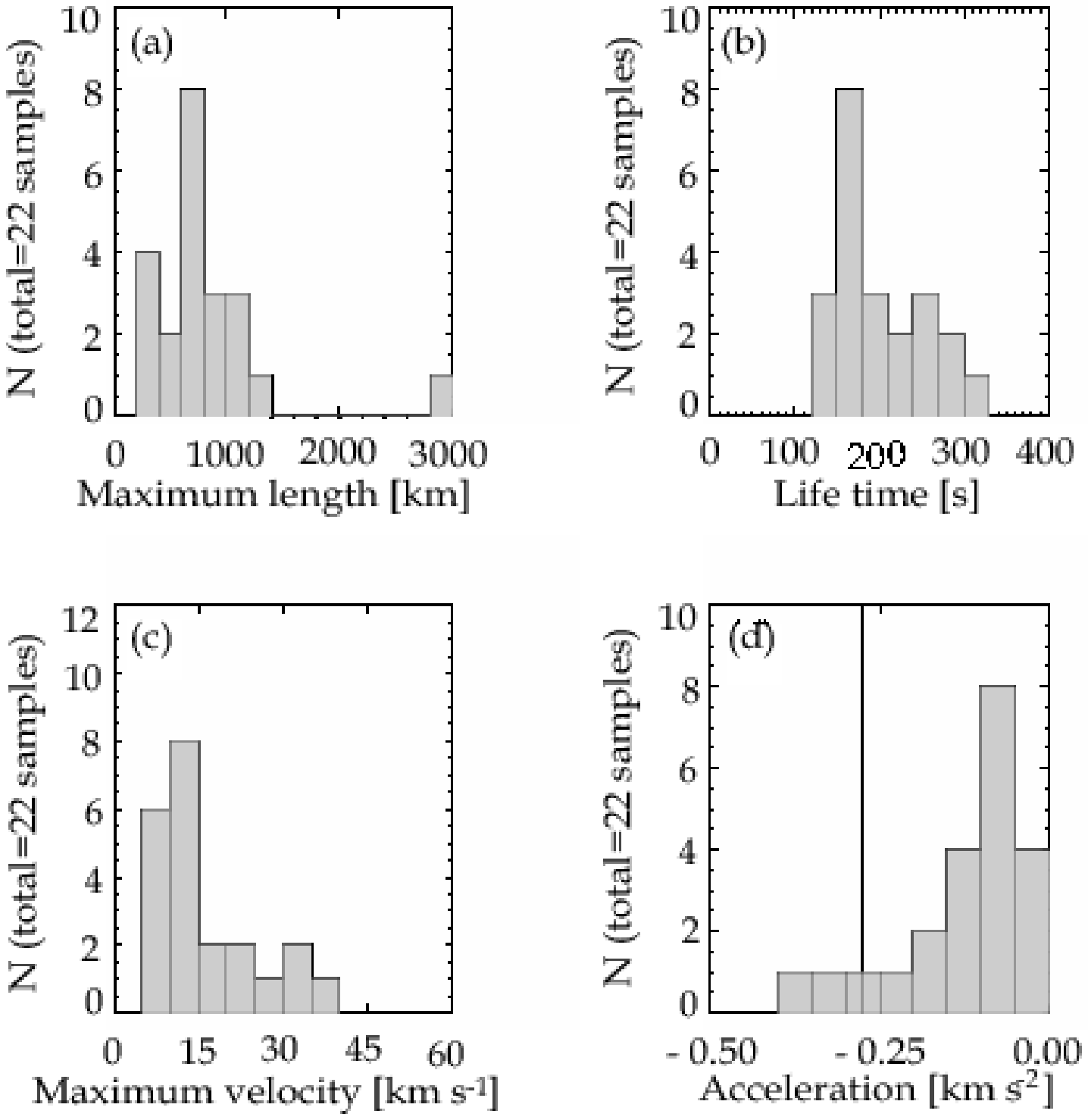}  
  \end{center}
  \caption{
  Histogram of (a) the maximum length, (b) the life time, (c) the maximum velocity and (d) the acceleration of 22 `fade out' spicules.
  The mean value of each spicule's maximum length, life time, maximum velocity and acceleration is 820\,km, 197\,s, 15.9\,km s$^{-1}$ and $-$0.13\,km s$^{-2}$, respectively.     
  The solid line in the histogram of acceleration indicate the solar surface gravity.     
  }\label{fig:periplot}
\end{figure}


\section{Summarizing Discussion}

In our observational work, we found the following: (1) In a plage region, we clearly identified spicular jet features. (2) They were shorter in length than quiet region limb spicules, and followed ballistic motion under constant deceleration. (3) Majority (80\%) of the plage spicules showed a full rise and retreat cycle, while 10\% of them faded out without a complete retreat. (4) The deceleration of the spicule was proportional to the maximum velocity at ejection ( i.e. the initial velocities ).

The dynamical characteristics of plage spicules found in this work, can be explained by the `shock-driven' models of spicules presented by Shibata \& Suematsu (1982), Hansteen et  al. (2006) and Heggland et al. (2007), although some extensions or modifications of their results must be taken into account to match the theory accurately with the observational data. 
For example, the lengths of the plage spicules were smaller than limb spicules, as predicted by Shibata \& Suematsu (1982), although it was necessary to change the height at which the pressure enhancements were excited to match the observed values. 
De Pontieu et al. (2004, 2007a) proposed that p-mode oscillations in the photosphere, instead of a single pulse, could drive the spicules and explain the behavior of dynamic fibrils.
Based on this result, Hansteen et al. (2006) and Heggland et al. (2007) showed that there exists a linear positive relation between the decelerations and the maximum velocity through studying the action of a train of N-shaped shock waves. 
We found that the relation itself held for plage spicules.
In our study, however, examples were found with greater velocities and greater decelerations than the parameter range simulated in their study. 

The origin of the `fade out' spicules remains unanswered in this study. 
We are yet to discover why the two types of spicules show different behavior in their retreat phase, even though their dynamic parameters are governed by the same relationship.
We have studied the two types to find any potential differences in, for example, their spatial distribution or the Ca\emissiontype{II} H brightness of their foot points. 
However no such relationship was found.
A difference in the thermal properties of the jets, such as heating due to thermal conduction or extra heating agencies may be the key to this problem. 

In conclusion, we found a large sample of plage spicules, and we derived their statistical properties.
These statistical properties showed that the spicules follow ballistic motion.
We think that those observed characteristics of plage spicules are well explained theoretically by the shock-driven model, when set in an atmospheres that is more realistic than considered thus far.

\begin{figure}
  \begin{center}
    \FigureFile(80mm,80mm){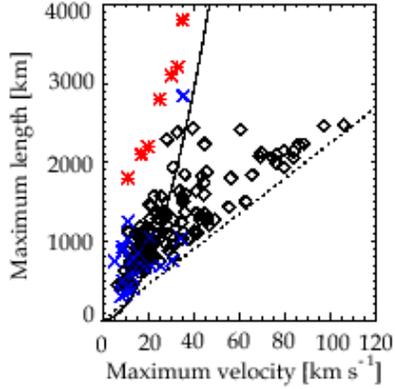}
  \end{center}
  \caption{
  Scatter plot of spicule's maximum length vs. its maximum velocity, with black diamonds for the `parabolic' spicules, with blue crosses for the `fade out' spicules, and with red asterisks for the simulation results done by Shibata \& Suematsu (1982).
   The solid line indicate the relation between the maximum length and its maximum velocity when the acceleration is the solar surface gravity.
   Due to our exclusion of short-lived spicules, data points will not appear below the dotted line.    
  }\label{fig:periplot}
\end{figure}

\begin{figure}
  \begin{center}
    \FigureFile(80mm,80mm){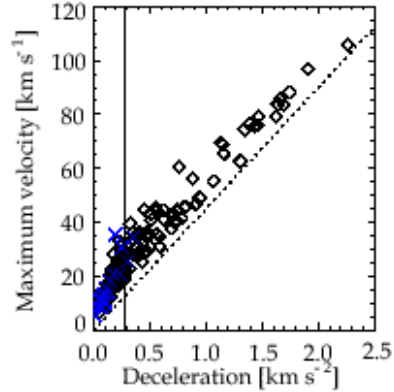}
  \end{center}
  \caption{
  Scatter plot of the deceleration vs. maximum velocity, with black diamonds for the `parabolic' spicules and with blue crosses for the `fade out' spicules.
  A solid line indicates the solar surface gravity.     
   Due to our exclusion of short-lived spicules, data points will not appear below the dotted line.    
    }\label{fig:periplot}
\end{figure}

\bigskip
We appreciate very much the comments by the anonymous referee, which helped us to clarify the contents of this paper.
The authors acknowledge all the staff and students of Kwasan and Hida Observatory whose comments were valuable for the improvement of the paper. The authors are supported by a grant-in-aid for the Global COE program "The Next Generation of Physics, Spun from Universality and Emergence" from the Ministry of Education, Culture, Sports, Science and Technology (MEXT) of Japan, and by the grant-in-aid for 'Creative Scientific Research The Basic Study of Space Weather Prediction' (17GS0208, PI: K. Shibata) from the Ministry of Education, Science, Sports, Technology, and Culture of Japan, and also partly supported by the grant-in-aid from the Japanese Ministry of Education, Culture, Sports, Science and Technology (No.19540474).  Hinode is a Japanese mission developed and launched by ISAS/JAXA, with NAOJ as domestic partner and NASA and STFC (UK) as international partners. It is operated by these agencies in co-operation with ESA and NSC (Norway). 


\end{document}